\documentclass[10pt]{article}

\usepackage[pdftex]{graphicx}
\graphicspath{{./pdf/}}

\usepackage{csagh}

\begin{document}
\begin{opening}

Computer Science 25(1) (2024) 25-46 \quad \quad \quad \quad \quad \quad https://doi.org/10.7494/csci.2024.25.1.5690

\title{Machine learning based event reconstruction for the MUonE experiment}

\author[The Henryk Niewodniczanski Institute of Nuclear Physics Polish Academy of Sciences, \URL{https://www.ifj.edu.pl}, e-mail: \URL{milosz.zdybal@ifj.edu.pl}]{Miłosz Zdybał}
\author[The Henryk Niewodniczanski Institute of Nuclear Physics Polish Academy of Sciences, \URL{https://www.ifj.edu.pl}, e-mail: \URL{marcin.kucharczyk@ifj.edu.pl}]{Marcin Kucharczyk}
\author[The Henryk Niewodniczanski Institute of Nuclear Physics Polish Academy of Sciences, \URL{https://www.ifj.edu.pl}, e-mail: \URL{marcin.wolter@ifj.edu.pl}]{Marcin Wolter}

\begin{abstract}
A proof-of-concept solution based on the machine learning techniques has been implemented and tested within the MUonE experiment designed to search for New Physics in the sector of anomalous magnetic moment of a muon. The results of the DNN based algorithm are comparable to the classical reconstruction, reducing enormously the execution time for the pattern recognition phase. The present implementation meets the conditions of classical reconstruction, providing an advantageous basis for further studies.
\end{abstract}


\end{opening}

\section{Introduction}

\begin{figure}[htbp]
    \begin{center}
        \includegraphics[width=1.0\textwidth]{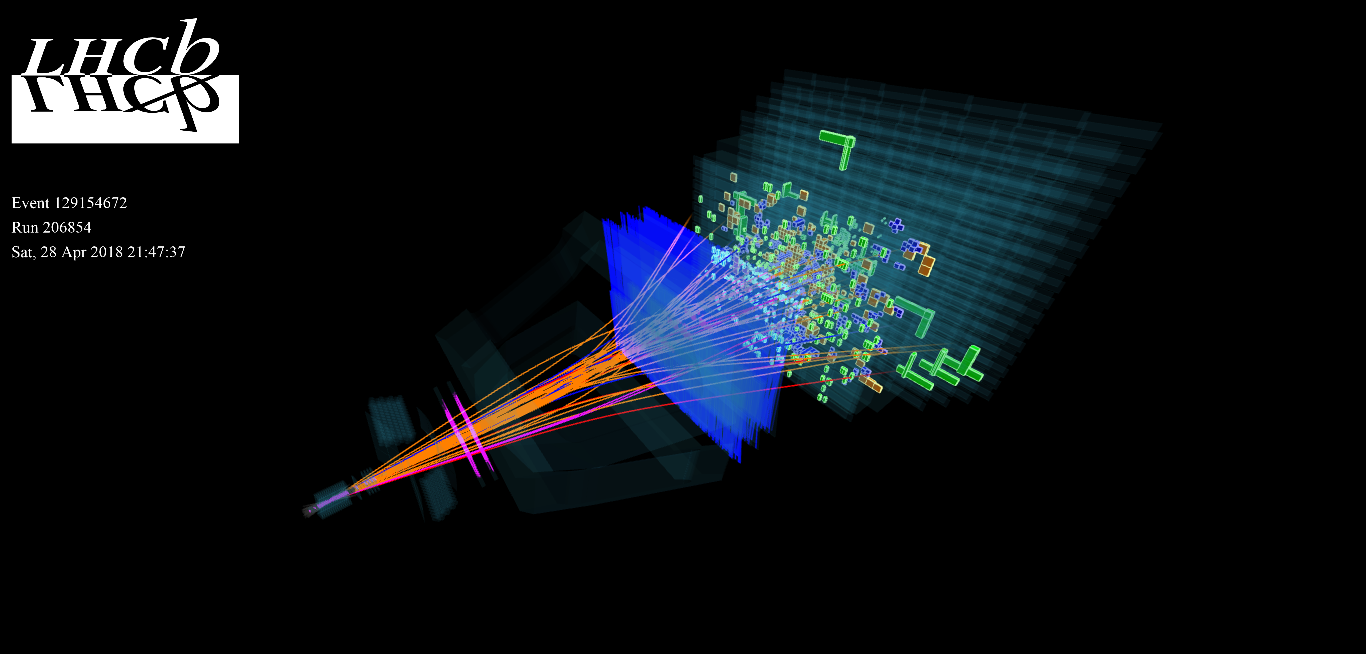} 
        \caption{Example of an event in High Energy Physics experiment, showing tracks of multiple particles passing through the detector~\cite{LHCbExperiment:2315673}. }
        \label{fig:LHCbEvent}
    \end{center}
\end{figure}

Significant developments have been applied during the last decades in the field of High Energy Physics (HEP) experiments, including computing technologies. Searches for New Physics phenomena, being an expansion of the so-called Standard Model, i.e. current incomplete theoretical knowledge about the basic behavior of the fundamental constituents of nature and the interactions between them, lead to experimental studies carried out at ever increasing energies. The number of particles created by the interaction of two particles (collision event) is generally increasing with the collision energy. As a consequence a huge number of charged particles have to be reconstructed (e.g. in proton-proton collisions), resulting in much more complex event patterns. A typical event in proton-proton collision showing the tracks of multiple particles passing through the detector is presented in Fig.~\ref{fig:LHCbEvent}, where the particles leave energy deposits (hits) in consecutive detector layers, being a basis for further track reconstruction. In order to enable the search for rare interesting collision events immersed in a huge background of events exhibiting well-known physics, the data rates related to the detector luminosity\footnote{Luminosity translates to the number of collisions per second and it is related to track density.} have increased enormously (e.g., 40 MHz readout rate in LHC). It has to be reduced online by more than five orders of magnitude before the information from an event is written on mass storage for further analysis.

This paper aims to review the machine learning based approach applied in crucial stages  of the data analysis process in HEP experiments, i.e. the procedure to determine basic kinematic parameters of charged particles at their point of production and the procedure to establish the location of these production points. They are commonly called track and vertex reconstruction. High density of tracks in a single collision event (detector occupancy) in operating and planned high-energy physics experiments results in a large combinatorics of hits in the event pattern recognition. Therefore, a novel machine learning based event reconstruction algorithms have been developed and tested within a framework of the MUonE experiment~\cite{MUonE} in order to maximize the statistical power of the final physics measurement. The results of the DNN based algorithm are comparable to the classical reconstruction, allowing not only to reduce execution time of the pattern recognition phase, but also to improve the precision and efficiency of the track and vertex reconstruction.

\section{Particle track reconstruction in High Energy Physics experiments}

\subsection{State of the art}

In High Energy Physics experiments the reconstruction of charged particle trajectories in the detectors is the most crucial process as it constitutes the major part of reconstruction time of the whole event. Tracking algorithms partition a collection of position measurements into groups corresponding to the to the hits originating from the same particle traversing through the detector. In the next step the parametrized trajectories are fitted to these collections to extract particle kinematics and locations of interaction vertices. The obtained  results are later combined with measurements from other detector systems, like calorimeters measuring the particle energy, to construct a complete physical model of  an event. In the last stage of the data analysis the reconstructed events are used  to extract the physical quantities.

\begin{figure}[htb]
    \begin{center}
        \includegraphics[width=0.9\textwidth]{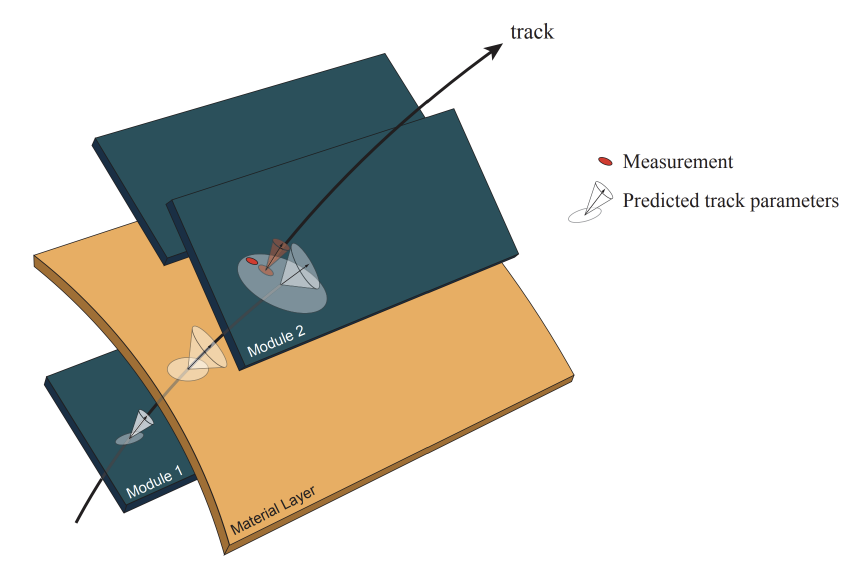} 
        \caption{Simplified illustration of a typical extrapolation process within a Kalman filter. The track representation on the detector module 1 is propagated onto the next measurement surface, which results in the track prediction on module 2~\cite{salzburger2007atlas}.
        }
        \label{fig:kalman}
    \end{center}
\end{figure}

Traditional tracking algorithms have been used with great success in the HEP experiments. However they suffer from serious  limitations that motivate for searching new solutions. These algorithms are inherently serial, and scale poorly with detector occupancy. In particle physics the measurement of  charged particle parameters is one of  the most computationally-intensive processes. This process relies on measurements of particle tracking detectors to construct a particle trajectory by combining the detected hits and resolving the particle momentum via fitting the trajectory points using the Kalman filter~\cite{kalman1960new}, The Kalman filter processes a set of discrete measurements to determine the internal state of a linear dynamical system (see Fig.~\ref{fig:kalman}). Both the measurements and the system can be subjected to independent random perturbations or noise. By combining predictions based on the previous state estimates with subsequent measurements, the impact of these perturbations on the
following state estimates can be minimized.

\begin{figure}[htbp]
    \begin{center}
        \includegraphics[width=0.9\textwidth]{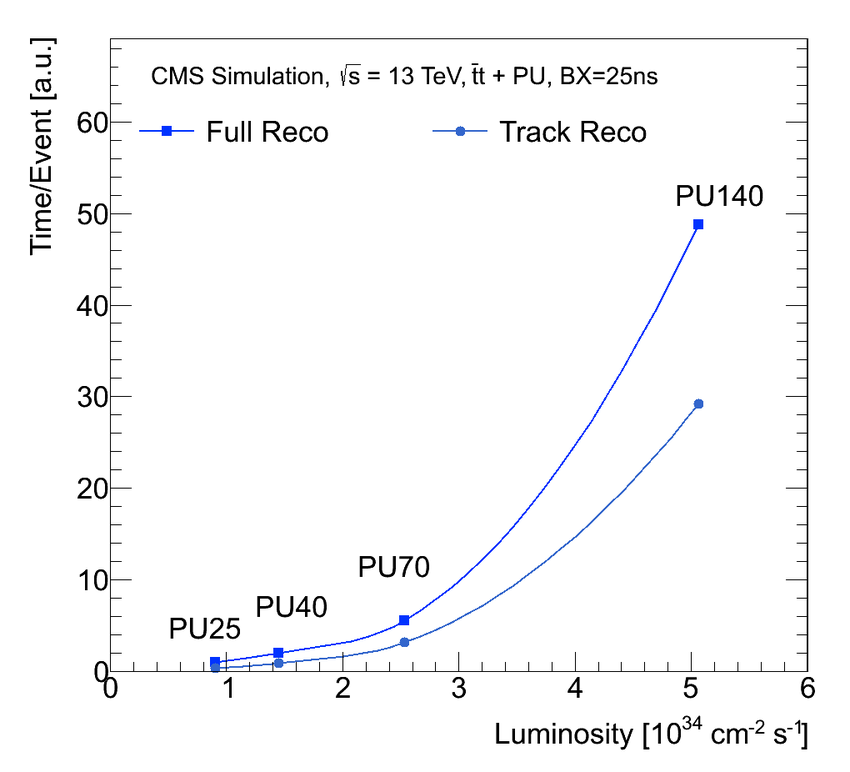} 
        \caption{Expected CPU time per event as a function of instantaneous luminosity collected by the CMS experiment, for both full reconstruction and the dominant tracking part. The pile-up (PU) is the number of interactions per beam crossing. PU25 corresponds to the data taken in 2012, and PU140 corresponds to the HL-LHC era. The CPU time of the reconstruction is dominated by the track reconstruction~\cite{cerati2017kalman}.}
        \label{fig:trackingCPU}
    \end{center}
\end{figure}

In high luminosity experiments, where multiple particles are produced as a result of an interaction, the process of isolating detector hits for each particle trajectory relies on considering each combination of hits that can potentially form a track and then fitting each hypothesis to determine which one represents a valid trajectory. Also the noise, always present in particle tracking detectors and resulting in additional hits in the detectors, increases the number of possible combinations. This process can be time-consuming, amounting to the most significant part  of the total data post-processing time. For such methods based on the Kalman filter the CPU needed for track reconstruction grows rapidly with the luminosity (see Fig.~\ref{fig:trackingCPU}). Therefore, more advanced methods of finding particle trajectories using the measurements from all active detector elements can be investigated. 

\subsection{Machine learning based track reconstruction}

Machine learning methods such as deep neural networks have some promising characteristics that could prove effective for particle tracking. Neural networks are known to be very good at finding patterns and modeling non-linear dependencies in data. They also involve highly regular computation that can run effectively on parallel architectures such as GPUs (Graphics Processing Units).

Neural Networks are therefore widely used in High Energy Physics not only as a classification tool, but also for other tasks like intelligent data reduction and time series analysis~\cite{kopciewicz2020upgrade} or reconstruction of a pulse shape from the front-end electronics~\cite{kopciewicz2021simulation}. 
They are used as well to optimize processes in various environments (Reinforcement Learning), for example automate the management of resources in a computing cloud~\cite{funika2023continuous}. 

The first approach to use neural networks for particle track reconstruction was done already in the 1980s~\cite{peterson1989track}. However modern techniques based on  deep learning have started to be studied in the last years. Two categories of  machine learning solutions, image-based and point-based models were investigated~\cite{farrell2017particle}.

The Convolutional Neural Networks (CNNs) (see for example  \cite{chollet2021deep} for description of various types of neural networks) proved to be a very efficient tool in image recognition. As such they are widely used in physics research, one of the examples is the CREDO experiment offline trigger using CNN to tag artefacts appearing in the CREDO database~\cite{piekarczyk2021article}. 

The computer vision techniques based on CNN such as semantic segmentation and image captioning have inspired the image-based models of particle track recognition. 
In this approach the detector data is treated as an image and the convolutional and also  recurrent neural networks are applied  to detect tracks. 

\subsubsection*{Image based track reconstruction}
\begin{figure}[htbp]
    \begin{center}
        \includegraphics[width=1.\textwidth]{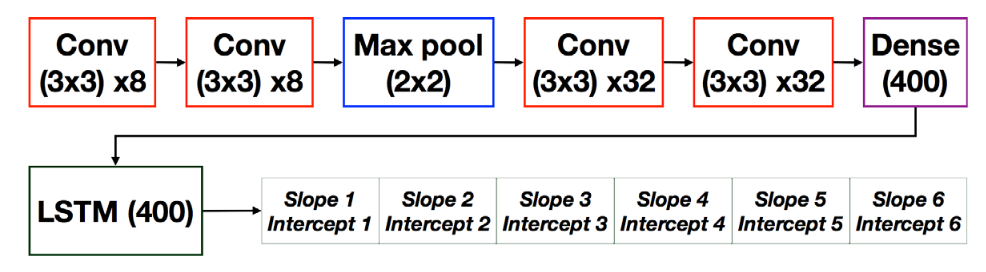} 
        \caption{Convolutional deep neural network with convolutional layers followed by dense and LSTM layers. Network is trained to reconstruct track parameters for multiple track events~\cite{Kucharczyk_Wolter_2019}.}
        \label{fig:cnnlstm}
    \end{center}
\end{figure}

In the image-based approach multiple track finding problem might be treated in a similar fashion to the image captioning, where the descriptions of the tracks (i.e., track parameters) are analogous to the text captions assigned to the various patterns seen in the image~\cite{vinyals2015show}. For this purpose, the long/short-term memory (LSTM)~\cite{hochreiter1997long} layer is used.  In the case of track reconstruction, the Convolutional Neural Network (CNN) sequentially reconstructs consecutive tracks, which are the sequential input for the LSTM layer. Example of such a network suited to reconstruct events with multiple tracks is shown in Fig.~\ref{fig:cnnlstm}.

The image-based models map nicely onto well-studied problems in computer vision and sequence modeling. However,  when scaling up to the realistic complexity of particle physics experiments they are suffering from high dimensionality and sparsity.

\subsubsection*{Point based track reconstruction}
In the point-based models, the  continuously distributed  spacepoint hits are used.  They are structured in a list or tree for learning how to group them into track candidates. A recurrent neural network acts as an iterative filter similar to a Kalman Filter. Therefore the model predicts the position of the point on the next detector layer. It can be used to build tracks by selecting the closest spacepoint to the prediction at every layer and searches until a complete track is found. This architecture might use an LSTM layer and  fully connected  layer with linear activation function to reconstruct multiple tracks.

\subsubsection*{Graph Neural Network models}
\label{GNNtracking}

In 2020 the Exa.TrkX project~\cite{ExaTrkX} has demonstrated the applicability of Geometric Deep Learning (GDL) methods to particle tracking~\cite{choma2020track} (specifically  Graph Neural Networks (GNNs)~\cite{scarselli2008graph}). GNNs have already proven to be succesfull in computer vision applications~\cite{krzywda2022graph}. Such a network is concerned with learning representations of data that have complex geometrical relationships and no natural ordering, which corresponds well with hits in the detector. In addition such models  are  naturally parallel and therefore well-suited to run on hardware accelerators and GPUs. The training of such a network might be computational demanding, but an answer of a trained network is fast and the computer time needed increases linearly with the number of tracks.

In  applications to track finding  graphs are constructed from the cloud of hits in each event. Edges are drawn between hits that may come from the same particle track according to some loose heuristic criteria. The GNN model is then trained to classify the graph edges as real or fake, giving a pure and efficient sample of track segments which can be used to construct full track candidates. Advanced studies concerning the application of GNNs for track reconstructions were presented by both CMS~\cite{arxiv.2203.01189} and ATLAS~\cite{caillou2022atlas} experiments.

It was shown~\cite{ju2021performance}, that within the simplifying assumptions, the GNN based track finding algorithm  can meet the tracking performance requirements of current, high luminosity collider experiments. This performance should be robust against systematic effects like detector noise, misalignment, and pile-up. The GNN based algorithms are promising and growing in popularity.

\section{MUonE experiment}

\begin{figure}[htbp]
    \begin{center}
        \includegraphics[width=0.7\textwidth]{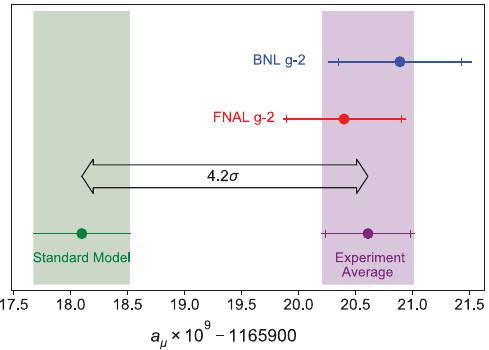} 
        \caption{Comparison of the measurements of anomalous muon magnetic moment $a_\mu$ with the Standard Model prediction \cite{PhysRevLett.126.141801}, where the discrepancy of 4.2 standard deviation between theory and experiment can be observed. In order to improve the visibility of the discrepancy the unit on the $x$-axis corresponds to a given $a_\mu$ value multiplied by $10^{9}$ and subtracted with 1165900.}
        \label{fig:MUonE_42sigma}
    \end{center}
\end{figure}

A very promising opportunity to search for New Physics in the sector of muon’s anomalous magnetic moment $a_\mu$ has appeared with the new results from $g-2$  experiments~\cite{PhysRevLett.126.141801,bennett2006final}, which measured the anomaly with respect to the Standard Model prediction at the level of 4.2 standard deviations (see Fig.~\ref{fig:MUonE_42sigma}). 

As the main limitation of eventual discovery comes from the precision of the theoretical Standard Model predictions, dominated by the uncertainty related to the hadronic interactions, the idea is to use the process of elastic muon scattering on electrons for the precise estimation of the hadronic contribution to $a_\mu$. The experiment dedicated to measure precisely such a hadronic contribution is the MUonE project~\cite{MUonE}, designed to determine the hadronic part of the running of the electromagnetic coupling constant in the space-like region by the scattering of high-energy muons on atomic electrons in a low-$Z$ target through the elastic process $\mu e \rightarrow \mu e$~\cite{Abbiendi_2017}. A result with significantly suppressed statistical uncertainty can be achieved on the hadronic contribution to $a_\mu$, which, together with the results from running~\cite{PhysRevLett.126.141801} or planned~\cite{abe2019new} $g-2$ experiments supposed to measure directly $a_\mu$, would increase the significance of observed discrepancy up to the level of 7 standard deviations. In order to measure the hadronic contribution with a required accuracy, a significant boost in precision and event statistics is necessary. This can only be achieved by accurate performance of the both the trigger and tracking system of the MUonE experiment.

\subsection{Experimental setup}
\label{sec:experimentalsetup}

The data samples of $\mu e \rightarrow \mu e$ elastic scattering will be collected in MUonE experiment using 150-160 GeV muons impinging on the atomic electrons of Beryllium targets. The upgraded M2 muon beam at the CERN SPS~\cite{Doble:250676} will be used for this purpose, delivering high energy and high-intensity muon and hadron beams, and also low intensity electron beams for calibration. The beams are conducted in the following way. First, the SPS primary proton beam of 450 GeV impinges on a primary Beryllium production target, where mainly secondary protons, electrons, pions and kaons are produced. In the next step the secondary particles are transported in a beam line allowing the pions and kaons to decay into muons. At this stage a 9.9~m thick Beryllium absorber stops the left-over hadrons, allowing the muons at the same time to pass basically unharmed. Next, such muons are momentum-selected employing large magnetic collimators, and finally the muons with momenta in the range of 100 and 225 GeV/c are selected. The typical maximal intensity for a beam energy of ~160 GeV is 5 $\times$ 10$^7$ $\mu$/sec.  

\begin{figure}
    \begin{center}
        \includegraphics[width=\textwidth]{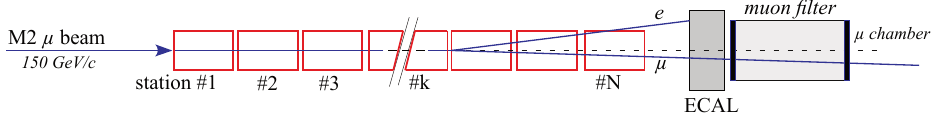}
        \caption{Schematic view of the MUonE experimental apparatus \cite{MUonE}.}
        \label{fig:MUonE_schematic1} 
    \end{center}
\end{figure}

\begin{figure}
    \begin{center}
        \includegraphics[width=\textwidth]{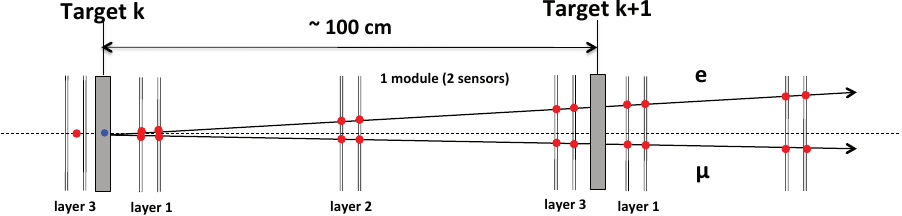} 
        \caption{Schematic view of a single tracking station~\cite{MUonE}.}
        \label{fig:MUonE_schematic2}
    \end{center}
\end{figure}

The main detectors of the MUonE experiment~\cite{MUonE} are specified in Fig.~\ref{fig:MUonE_schematic1} and \ref{fig:MUonE_schematic2}. The tracking system will provide the precise measurement of the scattering angles of the outgoing electron and muon, with respect to the direction of the incoming muon beam. It will contain 40 identical stations (see Fig.~\ref{fig:MUonE_schematic1}), each consisting of a 3~cm thick layer of Beryllium coupled to 3 Si layers (see
Fig.~\ref{fig:MUonE_schematic2}) located at a relative distance of about one meter from each other and spaced by intermediate air gaps. Such an arrangement provides both a distributed target with low-$Z$ and the tracking system. The silicon strip sensors for the MUonE project are characterized by a large active area sufficient to cover the full MUonE required acceptance, together with appropriate spacial resolution. They can also support the high readout rate of 40 MHz required for MUonE with their accompanying front-end electronics. The downstream particle identifiers are planned to be installed, required to solve the muon-electron ambiguity. That will be a calorimeter for the electrons and a muon filter for the muons. A homogeneous electromagnetic calorimeter placed downstream all the tracker stations will be used, in order to accomplish the physical requirements, i.e. particle identification, measurement of the electron energy and event selection.

\subsection{MUonE test beam in 2018}
\label{sec:2018testbeam}

\begin{figure}[htb]
    \begin{center}
        \includegraphics[width=0.7\textwidth]{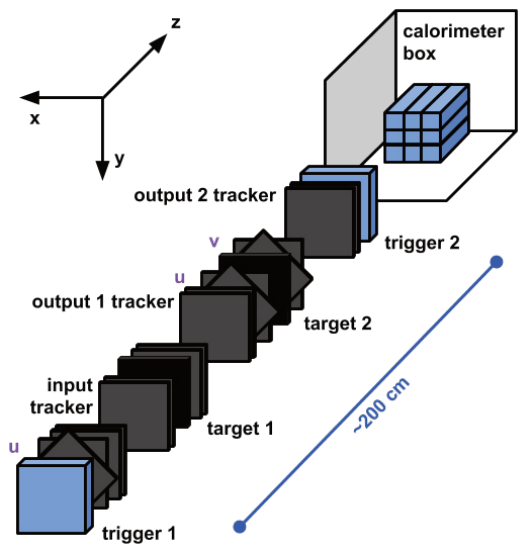}
        \caption{Schematic view of the apparatus used in the MUonE 2018 test run \cite{Abbiendi_2021}.}
        \label{fig:MUonE_MUonE_testbeamconfig} 
    \end{center}
\end{figure}

In 2018 the MUonE test run was performed with the aim to provide information for the design of the final MUonE detector setup~\cite{Abbiendi_2021}. The apparatus used was situated in the EHN2 experimental area, located behind the COMPASS spectrometer. The 187 GeV positive muon beam was obtained from decays of pions, which were stopped in the beam dump in the posterior part of the spectrometer. A 10 $\times$ 10 cm$^2$, 8~mm thick graphite target was followed by the tracking system with 16 microstrip layers consisted of a 9.293 $\times$ 9.293 $\times$ 0.041 cm$^3$ single-side sensor with 384 channels (see Fig.~\ref{fig:MUonE_MUonE_testbeamconfig}). Each tracking layer measured one hit coordinate, $x$ or $y$. Despite that, stereo stations rotated by an
angle $\pm \pi / 4$ were added. A calorimeter located at the end of the system, composed from BGO tapered crystals, covered an angular acceptance of about 15 mrad on each side from the center of Si layers.

From the data collected at the last period of the 6 months run, after the final requirements on the presence of an incoming track and at least two outgoing tracks, the number of events used in the analysis was reduced to ~94 $\times$ 10$^3$. The event reconstruction consisted of the following main steps: the pattern recognition, 2-dimensional track finding, combining 2-dimensional track candidates into a 3-dimensional track and finally, constructing a scattering event from three 3-dimensional tracks with a dedicated kinematic vertex fit based on a constrained least square method. Finally, this allowed to obtain a clean sample of $\mu e$ elastic scattering events.

The angular resolution was determined with the simulation, which met the MUonE 2018 test beam configuration. A sample of ${\sim 100 \times 10^3}$ events was produced, where the incoming muon beam was assumed to be a monoenergetic beam with energy of 187 GeV, and $x$ and $y$ distributions were adjusted to match the ones measured with data.

\section{Machine learning based track reconstruction for MUonE}

The measurement of the hadronic contribution to $a_\mu$ in the MUonE experiment requires novel fast and efficient real-time based algorithms for the track and vertex reconstruction, together with a flexible trigger system. New event reconstruction methods developed for the MUonE experiment, based on the novel hardware-triggerless techniques or, alternatively, on machine learning methods implemented on parallel GPU processing, may become a standard approach in the future High Energy Physics experiments, facing an enormously tight execution time imposed by a fully software trigger system and achieving a maximum possible event reconstruction efficiency and precision. It will allow to efficiently reduce the size of data expected to increase fastly in the future experiments, and also to maximize the statistical power of the final physics measurement. In the MUonE experiment the algorithms of track finding based on machine learning techniques are being developed and tested in order to speed up the reconstruction process. This may lead to the significant acceleration of the execution of pattern recognition algorithms in the real-time event reconstruction algorithms. Moreover, the use of DNN techniques may significantly improve both the reconstruction efficiency and the precision of measuring the parameters that are crucial for final measurement.

\subsection{Two-dimensional machine learning based reconstruction}

A first approach to use the machine learning techniques in context of the MUonE experiment~\cite{Kucharczyk_Wolter_2019} was based on an image-based model and a convolutional neural network~\cite{nature14539}.  This type of neural network is predominantly used in computer vision tasks, using a set of filters (called kernels) that analyze the image with a relatively small perception window (few pixels in size) that scans the input to produce the activation map of the filter. Network creates a set of filters sensible to different features in the input. Single filter can find multiple instances of the feature in the input image.

Training and testing datasets were generated using a two-dimensional toy-model. In total $40 \times 10^3$ events corresponding to an elastic scattering signal with the MUonE 2018 test beam configuration (see Sec.~\ref{sec:2018testbeam}) were produced. Each event contained one or two tracks reconstructed with the linear fit and was represented with a two-dimensional $28 \times 28$ pixel image. Optionally, the noise was also included.

The neural network was implemented in KERAS~\cite{chollet2015keras} with TensorFlow~\cite{abadi2016tensorflow} backend. It was trained to respond to the input image with slopes and intercepts of the tracks. Two convolutional layers were used with $3 \times 3$ convolution window, followed by the MaxPooling layer and another two convolutional layers. The dropout layer was used to control the overtraining. Final regression was performed using the 1024-node dense layer. The full network had over 2 million trainable parameters. For multi-track events, long/short-term memory (LSTM) mechanism was used. It was inspired by the work of HEP.TrkX project \cite{HEP.TrkX, arxiv.1810.06111}. To make events more realistic, noise and pixel inefficiency was introduced. Noise could be defined in the 0-30\% range, meaning a probability of pixel not belonging to the track to generate a signal. Pixel efficiency was lowered to 70\% by changing the probability of track pixel to generate a signal.

\begin{figure}
    \includegraphics[width=0.9\textwidth]{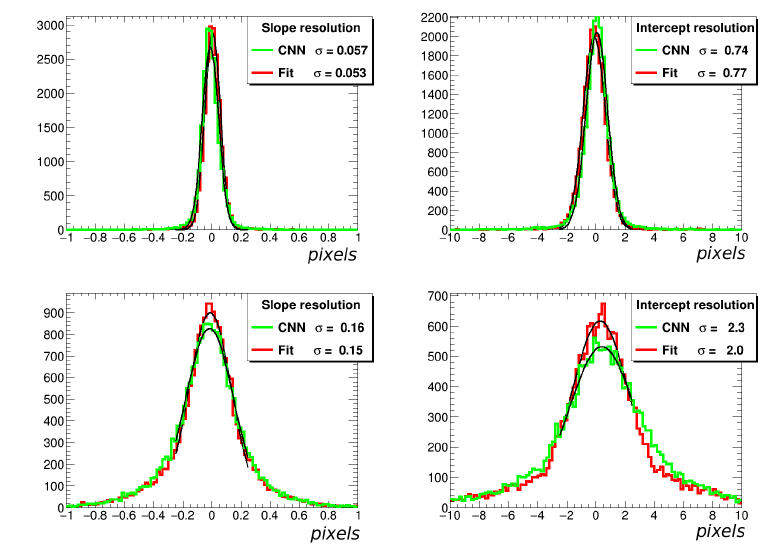}
    \caption{Comparison of the distributions of the difference between the reconstructed and true track parameters: slope (left) and intercept (right), between CNN-based and classical track reconstruction. Noise level at 10\% (top) and 30\% (bottom). Figure taken from~\cite{Kucharczyk_Wolter_2019}.}
    \label{fig:MUonE_2DfitResolution}
\end{figure}

Results provided by the CNN were used to find hits closest to the track candidates. The tracks were reconstructed using linear robust fit \cite{BRUN199781}. Differences between reconstructed and true tracks are shown in Fig.~\ref{fig:MUonE_2DfitResolution}, including 10\% and 30\% noise levels. Results were compared with the classical reconstruction algorithm (included in the plots). The neural network based approach proved to be successful and prompted the further development using three-dimensional approach.

\subsection{Three-dimensional DNN based track reconstruction}
\label{sec:MUonE_DNNreconstruction}

A natural next step after the work described in the previous section was an implementation of the machine learning based approach into three dimensions~\cite{10.1007/978-3-030-86887-1_21}, that would meet much better the requirements of the MUonE experiment. In general an artificial neural network was designed to reproduce properly the track parameters. Based on the set of hit coordinates, the network's task is to predict the slope and the intercept of the track for each of two outgoing \emph{\textmu-e} elastic scattering signal particles.

\subsubsection{Learning dataset}

The tracking detector of the MUonE experiment (in the final detector configuration as well as in the 2018 test run) is based on silicon strip sensors which provide only one value for the measurement that can be combined with the sensor position along the beam axis to create two-dimensional representation of the hit position, i.e. $(x, z)$ or $(y, z)$. An assumption was imposed that input data for the network will not include the information about the type of the hit ($x$, $y$, or $u$, $v$ for stereo layers), but all the hits will be ordered by an increasing $z$ coordinate.

The training dataset was generated using a leading-order event generator with the detector simulation preformed with \textsc{Geant4}~\cite{agostinelli2003geant4}. The sample  contained about $100 \times 10^3$ events corresponding to the MUonE 2018 test beam setup described in Sec.~\ref{sec:2018testbeam}. Input vector of the neural network consisted of 20 floating point values representing the measurements of the detector. Hits were arranged by the increasing $z$ coordinate, without distinction between $x$, $y$ and \emph{stereo} hits. The $z$ coordinates were not included in the input vector, as they were identical in all events. For each event a ground truth was provided in the form of slope and intercept of outgoing tracks, 12 values in total. The dataset was split into training and testing subsets in the 4:1 ratio.

\subsubsection{Artificial neural network}

The PyTorch~\cite{paszke2019pytorch} was chosen as the machine learning framework, as it incorporates tools needed for data handling, training process and inference, all with GPU support to accelerate underlying matrix-based computations. The input vector contained collection of hits described in the previous section. To reduce its size as well as the size of correlated network and its complicity, information not critical for the track reconstruction was removed. Hits were sorted by ascending $z$ value, which made explicit use of this coordinate, repeating in all events, redundant. In addition, hits related to the incoming muon were skipped, as the algorithm focused on the reconstruction of outgoing tracks. Final input vector included 20 values, each representing a measurement made by a silicon strip sensor. There was no distinction among $x$, $y$ and \emph{stereo} hits. The output vector contained slopes and intercepts of the two outgoing tracks, represented in \emph{x-z} and \emph{y-z} projections, totalling in 8 values. For ease of comparison, this is the same format as the ground truth was provided in.

\begin{figure}
    \includegraphics[width=0.7\textwidth]{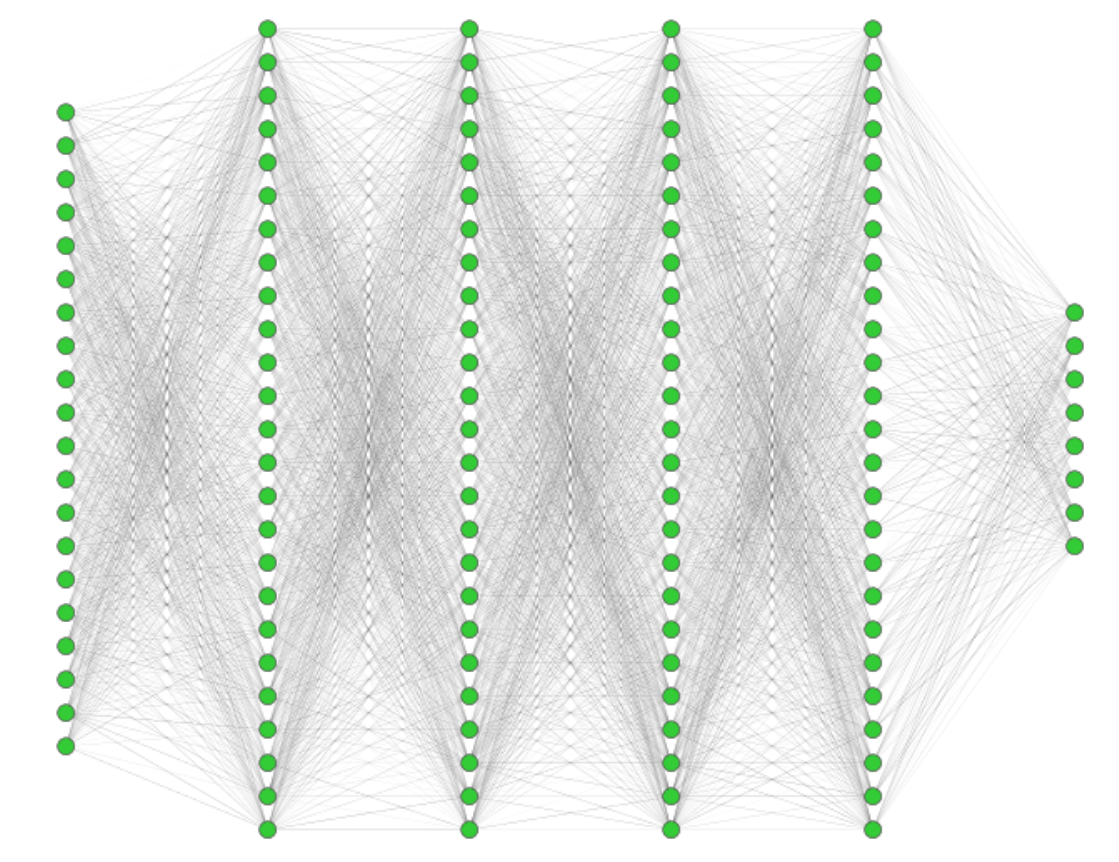}
    \caption{Architecture of the neural network used for track reconstruction. The input layer has 20 nodes, the following four hiddenlayers have 1000 nodes each and the output layer consists of 8 nodes. Number of nodes in the hidden layers in plot was reduced from 1000 to from 25 for clarity.}
    \label{fig:NNarch}
\end{figure}

The artificial neural network consisted of four fully connected layers of 1000 neurons each, with additional layers for input and output (see fig.~\ref{fig:NNarch}). It is important to mention that at this stage of development no hyperparameter optimization was performed, i.e. parameters of the network were arbitrarily set to achieve acceptable results in reasonable time during the development. The \emph{MSELoss} (Mean Squared Error Loss) \cite{mseloss} was chosen as the loss function, its  implementation from the PyTorch package was used. In the process of training, the network was optimized in a way that minimized the mean squared error between the output and ground truth. In terms of the activation function, ReLU (rectified linear unit) was used, as being suitable for deep neural networks.

\subsubsection{Reconstruction algorithm}
\label{sec:DNN_reconstructionalgorithm}

\begin{figure}
    \begin{center}
        \includegraphics[width=\textwidth]{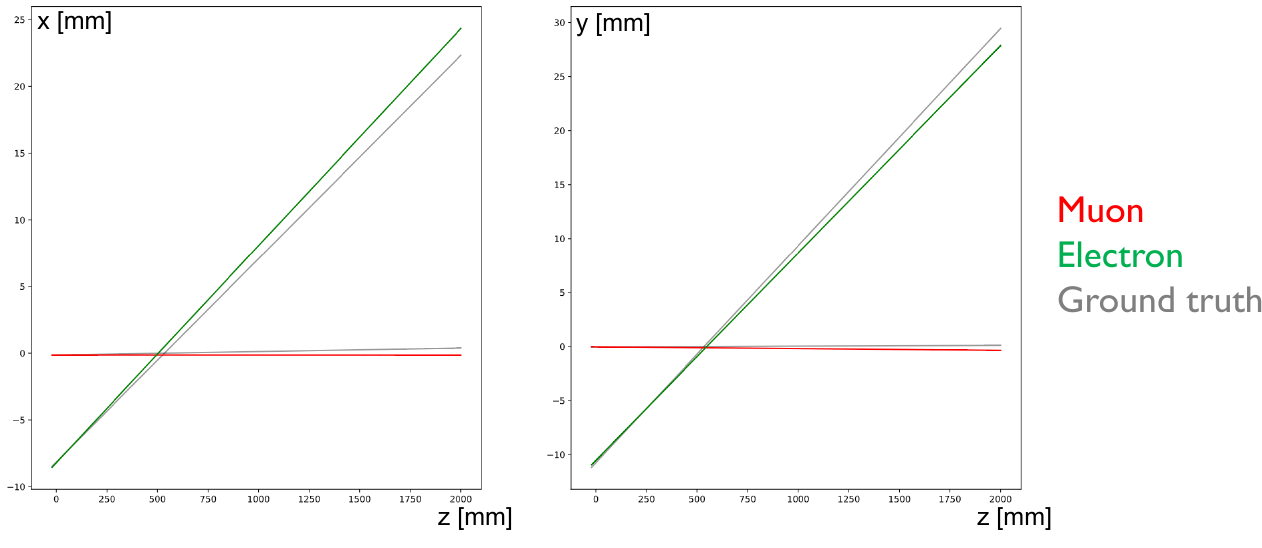}
        \caption{Example of the tracks reconstructed by the DNN, before applying further steps of the reconstruction algorithm. The ground truth shown in grey.}
        \label{fig:DNNexample} 
    \end{center}
\end{figure}

\begin{figure}
    \begin{center}
        \includegraphics[width=\textwidth]{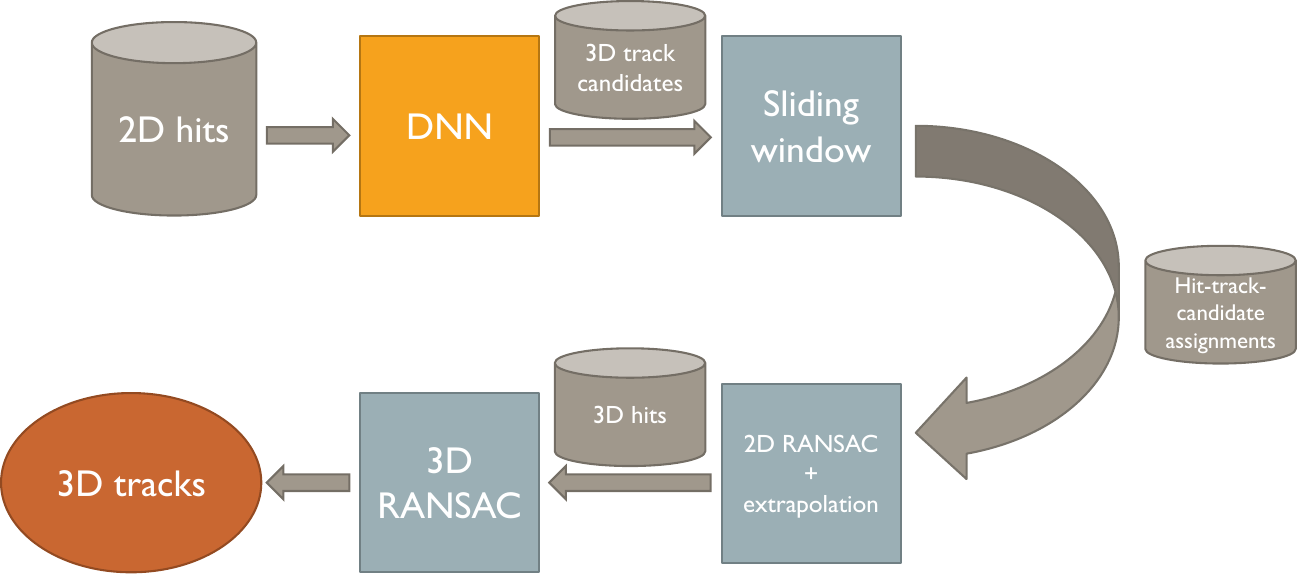} 
        \caption{The DNN-based algorithm for track reconstruction.}
        \label{fig:DNNalgo}
    \end{center}
\end{figure}

Comparison of the DNN-predicted tracks with the ground truth revealed that tracks are relatively close to each other (see Fig.~\ref{fig:DNNexample}), however reconstruction did not provide a precision required in the experiment. More complex algorithm had to be developed, where DNN was responsible only for the pattern recognition, being the most CPU time-consuming stage of the reconstruction in comparison to relatively fast linear fitting. Proposed algorithm is presented in Fig.~\ref{fig:DNNalgo} and described in the following sections.

\begin{itemize}

\item{Deep neural network based pattern recognition}

In the first step, the DNN machine learning model was used to turn the collection of hits representing \emph{\textmu-e} elastic scattering signal event into three-dimensional track candidates. Every hit was then assigned to the DNN-reconstructed track that was the closest geometrically in its plane. Example of the event with hits assigned to the tracks is shown in Fig.~\ref{fig:DNNalgo1}.

\begin{figure}
    \begin{center}
        \includegraphics[width=0.9\textwidth]{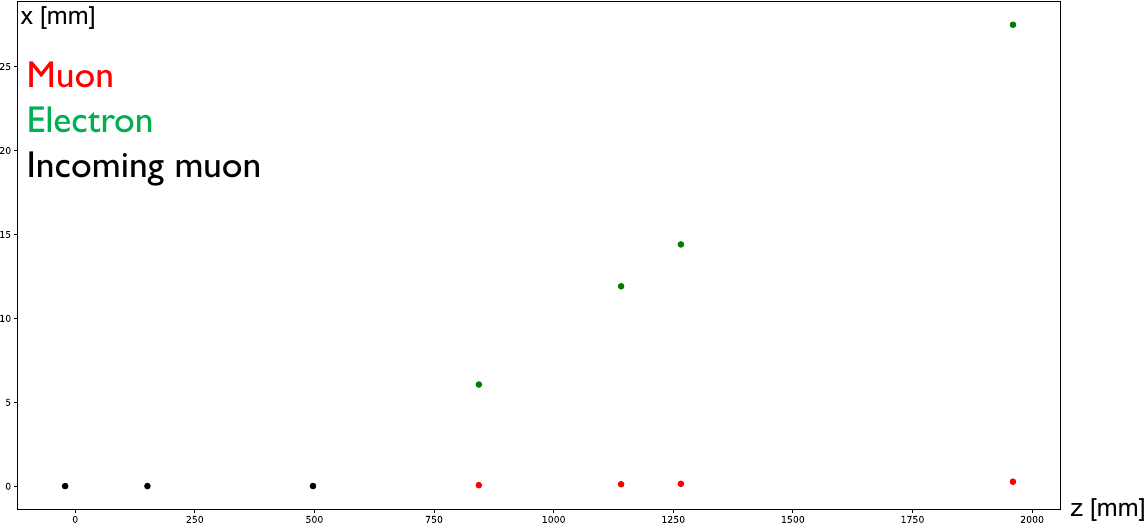}
        \caption{Example of the collections of hits corresponding to the \emph{\textmu-e} elastic scattering signal tracks constructed based on the DNN-predicted track candidates.
        Points represent the hits in \emph{x-z} projection, colours correspond to the particle type.}
        \label{fig:DNNalgo1} 
    \end{center}
\end{figure}

\item{Two-dimensional linear fit}

At this stage, RANSAC method (\emph{Random Sample Consensus}) \cite{10.1145/358669.358692} was used to reconstruct two-dimensional temporary tracks in \emph{z-x} and \emph{z-y} projections. The RANSAC iterative algorithm was chosen as it is effective for outlier removal. Implementation provided in Scikit-learn package \cite{scikit-learn} was used. As a result, two 2D lines were established in both \emph{z-x} and \emph{z-y} planes. Example of temporary tracks resulting from the 2D linear fit in \emph{z-x} projection is shown in Fig.~\ref{fig:DNNalgo2}.

\begin{figure}
    \begin{center}
        \includegraphics[width=\textwidth]{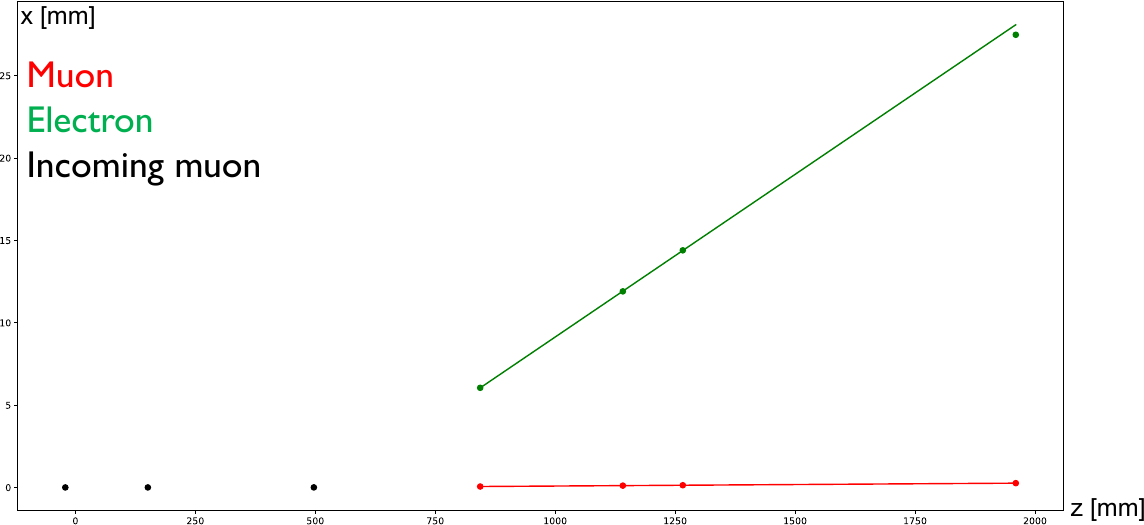} 
        \caption{Example of the result of the linear fit (solid lines) in \emph{x-z} projection for an outgoing muon and electron from the \emph{\textmu-e} elastic scattering signal event. The points represent the hits in \emph{x-z} projection, colours correspond to the particle type.}
        \label{fig:DNNalgo2}
    \end{center}
\end{figure}

\item{Final 3-dimensional track fit}

Two-dimensional temporary tracks from the previous step were used to extrapolate the missing coordinate for each hit. With a collection of 3D hits assigned to every track, the final linear fit was performed with 3-dimensional RANSAC algorithm. Example of reconstructed three-dimensional tracks is shown in Fig.~\ref{fig:DNNalgo3}.

\begin{figure}[htb]
    \begin{center}
        \includegraphics[width=0.9\textwidth]{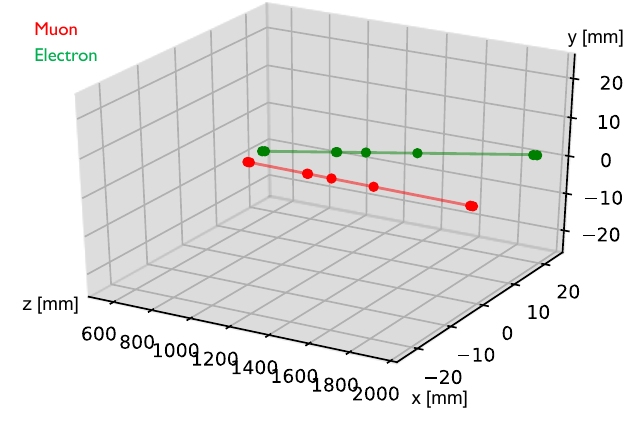} 
        \caption{Example of the result of the 3D linear track fit (solid lines - red for muon and green for electron) for an outgoing muon and electron from the \emph{\textmu-e} elastic scattering signal event. The points represent the hits, colours correspond to the particle type (red for
        muon and green for electron).}
        \label{fig:DNNalgo3}
    \end{center}
\end{figure}

\end{itemize}

\subsubsection{Results}

\begin{figure}[p]
    \begin{center}
        \includegraphics[width=\textwidth]{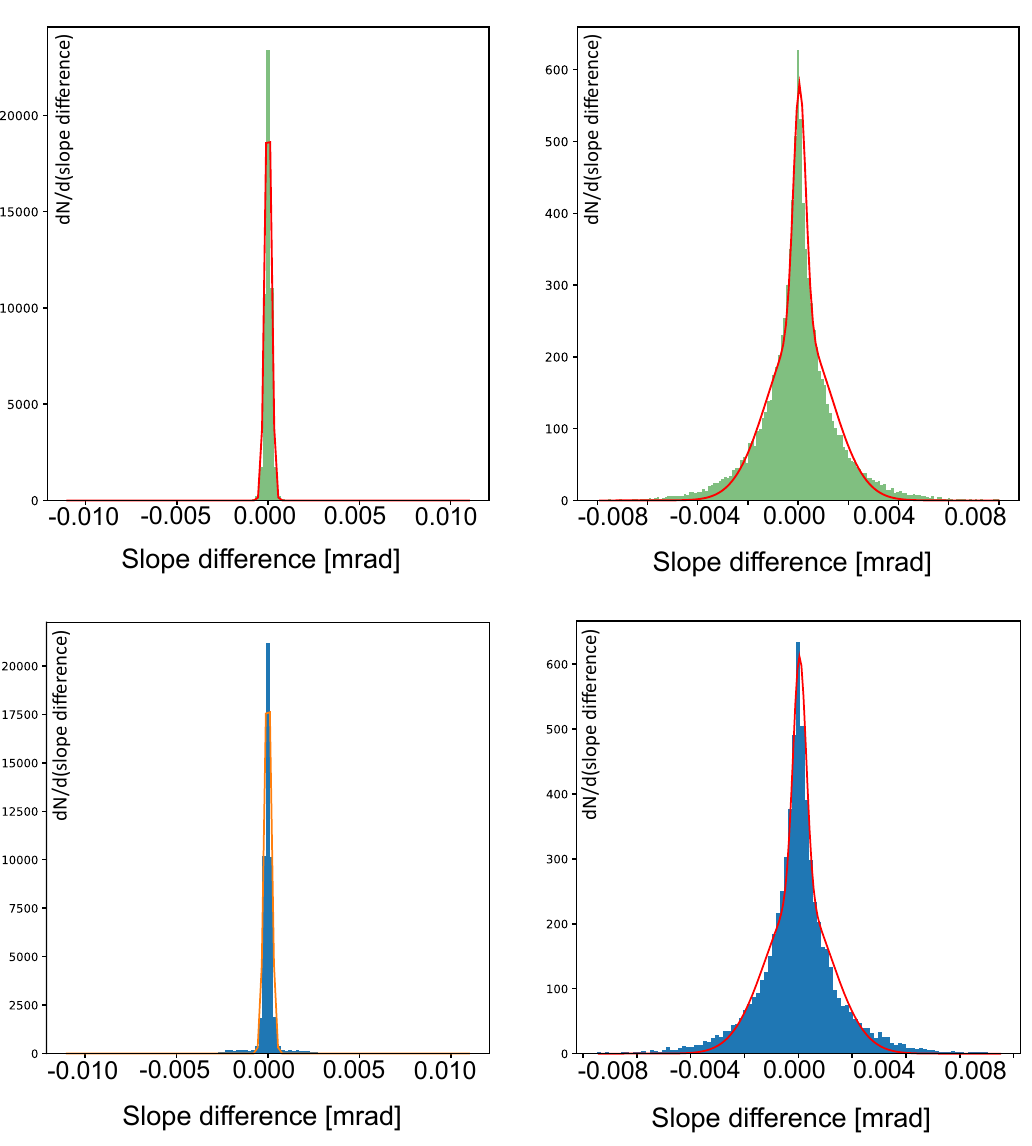} 
        \caption{Distributions of slope difference of reconstructed tracks (left for muons, right for electrons) in relation to the MC truth, for DNN-based algorithm (upper plots) and classical reconstruction (bottom plots).}
        \label{fig:DNNslopediff}
    \end{center}
\end{figure}

To assess the quality of the reconstructed tracks, the resolutions in track slopes were determined. To achieve this, histograms of difference in slopes between reconstructed track and ground truth were fitted with the Gaussian distribution, and the value of the standard deviation was interpreted as the resolution. In the case of electron track double Gaussian was used as this particle is more affected by the multiple scattering. The same procedure was performed for the tracks reconstructed with classical algorithm. Distributions of the slope differences are shown in Fig.~\ref{fig:DNNslopediff} and resolutions are summarized in Table~\ref{tab:DNNresolutions}. Additionally, efficiencies of the reconstruction algorithms were compared. Efficiency in this case is defined as the percentage of the tracks with the slope calculated with difference less than $1 \times 10^{-2}$ when compared to the ground truth. Efficiencies of reconstruction algorithms are compared in Table~\ref{tab:DNNefficiency}. Results achieved by the DNN-based algorithm are on pair with the conventional algorithm. Differences, if present, are not significant. This shows that the machine learning approach to track reconstruction has a great potential and should be investigated further.

\begin{table}
    \begin{center}
    \begin{tabular}{l|c|c}
        Particle & DNN based & Classical \\
        \hline\hline
        Muon & $\sigma = 0.000018$ mrad & $\sigma = 0.000019$ mrad \\
        \hline
        Electron & $\sigma_1 = 1.290$ mrad, & $\sigma_1 = 1.230$ mrad, \\
                 & $\sigma_2 = 0.245$ mrad  & $\sigma_2 = 0.244$ mrad \\
    \end{tabular}
    \end{center}
    \caption{\label{tab:DNNresolutions} Slope resolutions for an outgoing muon and electron. }    
\end{table}

\begin{table}
    \begin{center}
    \begin{tabular}{l|c|c}
        Particle & DNN based & Classical \\
        \hline\hline
        Muon & 100\% & 99.98\% \\
        \hline
        Electron & 99.66\% & 99.38\% \\
    \end{tabular}
    \end{center}
    \caption{\label{tab:DNNefficiency} Efficiencies of reconstruction algorithms, as defined in the text.}   
\end{table}

\section{Summary and outlook}

The present DNN based algorithm prototype for the three-dimensional track reconstruction in the MUonE experiment proved to be competitive with the classical track reconstruction tasks in terms of quality, with potential performance benefits. Further development will involve the implementation of the neural network architecture based on Graph Neural Network. Graph Neural Networks (GNNs, subsection~\ref{GNNtracking}) ensure, among others, the inductive bias, reduction of number of parameters, more elaborated loss function, and above all a much more natural data representation. New event reconstruction methods being developed for the MUonE experiment, based on the novel hardware-triggerless techniques using machine learning methods may become a standard approach in the future High Energy Physics experiments, facing an enormously tight execution time imposed by a fully real-time reconstruction and achieving a maximum possible efficiency and precision. Although such techniques are not yet applied on a scale in any High Energy Physics experiments, they are intensively developed and planned to be employed in the near future.

\begin{acknowledgements}
This research was supported by the National Science Centre NCN (Poland) under the contract no. 2022/45/B/ST2/00318 and also supported in part by PL-Grid Infrastructure.
\end{acknowledgements}

\bibliographystyle{cs-agh}
\bibliography{bibliography}

\end{document}